\documentclass{appolb}
\usepackage{graphicx}

\begin{document}
\title{Scalar Glueball in a Top-Down Holographic Approach to QCD%
\thanks{Presented at EEF70 -- Workshop on Unquenched Hadron Spectroscopy:
Non-Perturbative Models and Methods of QCD vs. Experiment at the occasion of Eef van Beveren's 70th birthday,
1-5 September 2014, University of Coimbra, Portugal}%
}
\author{Denis Parganlija
\address{Institut f\"{u}r Theoretische Physik, Technische Universit\"{a}t Wien, Wiedner Hauptstr.\ 8-10, 1040 Vienna, Austria}
}
\maketitle
\begin{abstract}
Identification of glueballs -- bound states of gauge bosons in Quantum Chromodynamics (QCD) -- is 
a very important open question in dynamics of the strong interaction. The search for the glueball ground state, carrying scalar quantum numbers, poses a 
particular challenge due to the existence of ({\it i}) several candidates for its realisation in the physical spectrum and
({\it ii}) inevitable mixing of the pure glueball state with those comprised of quarks. In this article, I discuss 
implications of an approach
in holographic QCD where, among others, the mass and the two-pion decay of the pure scalar glueball can be studied.
\end{abstract}
\PACS{11.25.Tq,12.38.Lg,12.39.Mk,13.25.Jx,14.40.Be}
  
\section{Introduction}

The non-Abelian nature of Quantum Chromodynamics (QCD) -- the theory of strong interaction --
gives rise to the expectation that its gauge bosons, the gluons, form composite objects
denoted as glueballs \cite{Salomone:1980sp}. These states would have access to various quantum numbers $J^{PC}$,
where $J$ denotes the total spin, $P$ the parity and $C$ the charge conjugation;
the corresponding spectrum in the QCD Yang-Mills sector has been determined in numerical simulations
\cite{Morningstar:1999rf,Gregory:2012hu} but the identification of glueballs in experimental data has proven to be a challenge,
particularly in the $J^{PC} = 0^{++}$ (scalar) channel.
\\\\
There are at least two reasons to study glueball states. Firstly, their masses are generated solely by the strong
interaction; no influence of the Higgs mechanism -- providing, e.g., quarks with a current mass -- is present
in the glueball mass generation. Secondly, due to their structure glueballs must possess integer spin assigning
them to mesons; our understanding of the physical meson spectrum would not be complete without glueballs.  
\\
\\
Glueball ground state is scalar \cite{West}; listings of the Particle Data Group (PDG) \cite{PDG} contain five isospin-zero scalar states in 
the energy region below 1 GeV: $f_0(500)$ or $\sigma$,  $f_0(980)$,  $f_0(1370)$,  $f_0(1500)$ and  $f_0(1710)$.
While all of them possess the quantum numbers of the ground state, there are strong reasons
to focus on resonances above 1 GeV as candidates for the scalar glueball:
($i$) simulations in lattice QCD determine the ground-state mass at between approximately 1.65 GeV and 1.8 GeV 
\cite{Morningstar:1999rf,Gregory:2012hu};
($ii$) various effective approaches to low-energy QCD arrive at an analogous result for the mass while
describing the overall dynamics correctly \cite{Close:2001ga}.
Complication is that $f_0$ resonances will most certainly have contributions not only from the (pure) glueball states but also
from those containing quarks ($\bar{q}q$ \cite{Parganlija:2010fz}, $\bar{q}\bar{q}qq$ \cite{Jaffe:1976ig} and others).
This leads to various issues in both theory and experiment \cite{Bugg:2004xu,Parganlija:2012nc}
and represents the main reason why the scalar glueball has still not been clearly identified.
\\
\\
In this article, the question of the scalar glueball is tackled by a holographic approach to non-perturbative
QCD.
Such approaches are based on the 
idea of 
a duality between strongly coupled quantum gauge
field theories and weakly coupled supergravity/superstring
theories in one dimension higher, pursuing Maldacena's
groundbreaking conjecture of a complete equivalence between the supergravity limit of type-IIB string theory on an 
$AdS_5 \times S^5$ space 
and the large-$N$ limit of an $\cal{N}$ $=4$ supersymmetric and conformal $U(N)$ gauge theory 
on its boundary (AdS/CFT correspondence \cite{Maldacena}).
In Ref.\ \cite{Witten}, Witten proposed an analogous duality in type-IIA string 
theory,
with supersymmetry and conformality not present in line with their 
absence in QCD.
The supersymmetry is broken by compactification on a circle ($S^1$);
for a vanishing circle radius, the 5-dimensional pure Yang-Mills theory is reduced to a 4-dimensional one.
However, the supergravity approximation
requires finite circle radius (whose inverse is defined as the
Kaluza-Klein mass $M_{\mbox{\tiny KK}}$), and also a large coupling.
Thus constructed holographic approaches are referred to as {\it top-down} models \cite{Klebanov2000,SaSu};
there are also more phenomenological {\it bottom-up}
constructions -- see Ref.\ \cite{Afonin:2010fr} and references therein.
This articles describes the implications of the top-down Witten-Sakai-Sugimoto
model \cite{SaSu} for the glueball spectroscopy; more details can be found in Ref.\ \cite{BPR}.
  
\section{The Model and Its Implications}

While Witten's model contained only gauge fields, the novel feature of Witten-Sakai-Sugimoto model is the
inclusion of chiral quarks introduced by $N_f$ (number of flavours) probe D8 and anti-D8 branes [inducing $U(N_f) \times U(N_f)$ chiral
symmetry] that extend along all dimensions
of the 10-dimensional space with the exception of a 
(Kaluza-Klein) circle. The branes are usually antipodally separated
with regard to this $S^1$. The space geometry, however, is such that the branes and antibranes merge at a certain point --
interpreted as realisation of the chiral-symmetry breaking.
\\\\
Up to a Chern-Simons term, the corresponding action for D8-branes reads

\begin{equation}
S_{\mbox{\tiny D8}} = -T_{\mbox{\tiny D8}} \, \mbox{Tr} \int \mbox{d}^9 x e^{-\Phi} \sqrt{- \mbox{det}(\tilde{g}_{MN} + 2\pi \alpha ' F_{MN})}
\end{equation}
where $T_{\mbox{\tiny D8}} = (2\pi)^{-8} l_s^{-9}$ (and $l_s^2 = \alpha '$, with $\alpha '$ the string 
coupling), $g_{MN}$ is the metric of the D-brane world volume, $\Phi$ is the dilaton field and $F_{MN}$ a field strength
tensor whose components are, upon dimensional reduction, identified as meson fields of interest. No backreaction
of the Witten-model background to D8-branes is considered; consequently $N_f$ is fixed and, as can be argued \cite{SaSu},
significantly smaller than the number of colours (large-$N_c$ limit).
\\\\
The above action is expanded up to the second order in fields as
\begin{equation}
S_{\mbox{\tiny D8}}^{(2)} = - \kappa \, \mbox{Tr} \int \mbox{d}^4 x \int_{-\infty }^{\infty} \mbox{d}Z
\left[ \frac{1}{2} \, K^{-\frac{1}{3}} \, \eta^{\mu \rho} \eta^{\nu \sigma} F_{\mu \nu}F_{\rho \sigma} + M_{\mbox{\tiny KK}}^2 \, \eta^{\mu \nu}
F_{\mu Z} F_{\nu Z} \right ]
\end{equation}
where $\kappa = \lambda N_c / (216 \pi^3)$ \cite{BPR}, $\lambda = g_{\mbox{\tiny YM}}^2 N_c$ is the 't Hooft coupling (and $g_{\mbox{\tiny YM}}$
the 4-dimensional coupling),
$Z$ is the holographic radial coordinate (and $K = 1+Z^2$) and
$\eta^{\mu \nu}$ is the flat metric diag$(-,+,+,+)$. 
The Kaluza-Klein mass $M_{\mbox{\tiny KK}}$ sets the model scale;
beside the scale, the model contains only one unknown
quantity: the coupling $\lambda$.
They are usually determined such that the mass of the rho meson
and the pion decay constant
correspond to their physical values. This yields $M_{\mbox{\tiny KK}} = 949$ MeV and $\lambda = 16.63$;
alternative methods for their determination do not alter
model conclusions \cite{BPR}.
\\\\
Once $M_{\mbox{\tiny KK}}$ and $\lambda$ are known, the model describes
various experimental quantities surprisingly well
\cite{SaSu}. Glueballs are of interest in this article;
building on the work of Ref.\ \cite{BMT},
the following masses in the scalar channel are obtained:
\begin{eqnarray}
M_{G_E} &=& 855 \mbox{ MeV;} \;\; M_{G^*_E} = 2168 \mbox{ MeV} \nonumber \\
M_{D_E} &=& 1487 \mbox{ MeV;} \;\; M_{D^*_E} = 2358 \mbox{ MeV.}
\end{eqnarray}
Two types of glueball states are present: one, denoted with $G_D$, is predominantly dilaton while the other,
denoted with $G_E$, 
involves a graviton
polarisation in a fourth spatial dimension, unlike
the dilaton, and has therefore been termed {\it exotic} \cite{CM}. Excited states are denoted with an asterisk. We observe that the exotic ground state
is approximately 50\% lighter than the expectation from lattice QCD; the dilaton mode mass is only approximately
15\% smaller than the lattice result. Numerical simulations also indicate the mass of the first excited scalar
state to be $\simeq$ 2600 MeV with errors amounting to $\simeq$ 300 MeV \cite{Morningstar:1999rf} (although 
mass corrections in the unquenched case may be substantial \cite{Gregory:2012hu}); the excited dilaton state is within errors consistent
with the lattice result while the excited exotic mode is approximately 15\% too light. Hence already from the mass results
the indication is that the exotic mode could be discarded while the dilaton mode appears compatible with simulations
of the Yang-Mills sector of QCD.
\\\\
This is corroborated by the decay ratios of the modes. 
The corresponding Lagrangians are obtained by
inserting 10-dimensional metric fluctuations (whose explicit forms together with further details
are presented in Ref.\ \cite{BPR})
into the action for D8 branes and integrating over the bulk coordinates (see also Ref.\ \cite{Hashimoto}); ratios of decay widths $\Gamma$ and the
respective masses $M$ read 
\begin{eqnarray}
\Gamma_{G_E \rightarrow \pi\pi}/M_{G_E} &=& 0.092; \;\; \Gamma_{G_E^* \rightarrow \pi\pi}/M_{G_E^*} = 0.149 \nonumber
\\
\Gamma_{G_D \rightarrow \pi\pi}/M_{G_D} &=& 0.009; \;\; \Gamma_{G_D^* \rightarrow \pi\pi}/M_{G_D^*} = 0.011.
\end{eqnarray}
The two main candidates for the scalar glueball are the $f_0(1500)$ and $f_0(1710)$ resonances 
(see Ref.\ \cite{Parganlija:2012nc} and references therein). Experimental data imply 
$\Gamma_{f_0(1500) \rightarrow \pi\pi}/M_{f_0(1500)} = 0.025 \pm 0.003$ \cite{PDG} and
$\Gamma_{f_0(1710) \rightarrow \pi\pi}/M_{f_0(1710)} \simeq 0.009$ -- $0.017$ \cite{PhD}.
The exotic mode is thus too broad, again indicating that its interpretation as a
physical state is uncertain; contrarily, the ratio $\Gamma/M$ for the dilaton mode
is within the experimental interval for $f_0(1710)$.\\
Similar is true for the excited states: 
$\Gamma_{G_E^* \rightarrow \pi\pi}$ is larger than $\Gamma_{G_D^* \rightarrow \pi\pi}$
and nicely comparable to widths of
$f_0$ states near and above 2 GeV; however, the full decay width of $G_E^*$ having contributions from $2K$, $2\eta$,
$4\pi$ and other channels is unphysically large ($\sim$ 1 GeV, see Ref.\ \cite{BPR}). Contrarily,
$\Gamma_{G_D^* \rightarrow \pi\pi}$ is small but the full decay width of $G_D^*$ is of the order of 460 MeV \cite{BPR}
and thus significantly closer to the data \cite{PDG} whose current uncertainties unfortunately do not allow for a clear identification of an excited scalar glueball.

\section{Summary and Outlook}
In this article, a top-down holographic approach to low-energy QCD -- Witten-Sakai-Sugimoto model -- has been presented and its 
implications for phenomenology of scalar glueballs have been discussed. The model offers two sets of glueball states, a dilaton and an
{\it exotic} mode that, unlike the dilaton,
involves a graviton polarisation in a fourth spatial dimension.
The exotic ground state has a mass approximately 50\% smaller than the value expected in lattice QCD; its $2\pi$ decay width
is substantially larger than that of the two prime candidates for the scalar glueball, the resonances $f_0(1500)$ and
$f_0(1710)$. Contrarily, the mass of the dilaton mode (= 1487 MeV) is quite close to masses of both mentioned resonances;
its $2\pi$ decay width is within the experimental range for $f_0(1710)$, which therefore appears to be the preferred
candidate for the glueball ground state.
In the excited channel,
the exotic state is unphysically broad while the dilaton 
width $\sim 460$ MeV is close to the (still ambiguous) data on $f_0$ states near/above 2 GeV.\\
Nonetheless, further pursuit of glueball dynamics in holography is called for, particularly in light of expectations
from the planned PANDA experiments at FAIR \cite{PANDA}.

\section{Acknowledgments}

I am grateful to F.~Br\"{u}nner and A.~Rebhan for collaboration and to D.~Bugg and S.~Janowski for extensive discussions.
I am also grateful to the Workshop Organisers for their generous support of my participation. This work
is supported by the Austrian Science Fund FWF, project no.\ P26366.


\begin{thebibliography}{99}

\bibitem{Salomone:1980sp} 
A.~Salomone, J.~Schechter and T.~Tudron,
Phys.\ Rev.\ D {\bf 23}, 1143 (1981);  
C.~Rosenzweig, A.~Salomone and J.~Schechter,
Nucl.\ Phys.\ B {\bf 206}, 12 (1982)  [Erratum-ibid.\ B {\bf 207}, 546 (1982)];  
A.~A.~Migdal and M.~A.~Shifman,
Phys.\ Lett.\ B {\bf 114}, 445 (1982);  
H.~Gomm and J.~Schechter,
Phys.\ Lett.\ B {\bf 158}, 449 (1985);  
R.~Gomm, P.~Jain, R.~Johnson and J.~Schechter,
Phys.\ Rev.\ D {\bf 33}, 801 (1986).  


\bibitem{Morningstar:1999rf} 
C.~J.~Morningstar and M.~J.~Peardon,
Phys.\ Rev.\ D {\bf 60}, 034509 (1999)  [hep-lat/9901004];  
Y.~Chen {\it et al.},
Phys.\ Rev.\ D {\bf 73}, 014516 (2006)  [hep-lat/0510074]; 
M.~Loan, X.~Q.~Luo and Z.~H.~Luo,
Int.\ J.\ Mod.\ Phys.\ A {\bf 21}, 2905 (2006)  [hep-lat/0503038].  

\bibitem{Gregory:2012hu} 
E.~Gregory, A.~Irving, B.~Lucini, C.~McNeile, A.~Rago, C.~Richards and E.~Rinaldi,
JHEP {\bf 1210}, 170 (2012)  [arXiv:1208.1858 [hep-lat]].  

\bibitem{West} 
  G.~B.~West,
  Phys.\ Rev.\ Lett.\  {\bf 77}, 2622 (1996)
  [hep-ph/9603316].

\bibitem{PDG} K.~A.~Olive \textit{et al}. (Particle Data Group), Chin.\ Phys.\ C, 
\textbf{38}, 090001 (2014).
  
  
\bibitem{Close:2001ga} 
F.~E.~Close and A.~Kirk,
Eur.\ Phys.\ J.\ C {\bf 21}, 531 (2001)  [hep-ph/0103173];  
F.~Giacosa, T.~Gutsche and A.~Faessler,
Phys.\ Rev.\ C {\bf 71}, 025202 (2005)  [hep-ph/0408085];  
F.~Giacosa, T.~Gutsche, V.~E.~Lyubovitskij and A.~Faessler,
Phys.\ Lett.\ B {\bf 622}, 277 (2005)  [hep-ph/0504033];  
F.~Giacosa, T.~Gutsche, V.~E.~Lyubovitskij and A.~Faessler,
Phys.\ Rev.\ D {\bf 72}, 094006 (2005)  [hep-ph/0509247];  
H.~Y.~Cheng, C.~K.~Chua and K.~F.~Liu,
Phys.\ Rev.\ D {\bf 74}, 094005 (2006)  [hep-ph/0607206];  
V.~Mathieu, N.~Kochelev and V.~Vento,
Int.\ J.\ Mod.\ Phys.\ E {\bf 18}, 1 (2009)  [arXiv:0810.4453 [hep-ph]];  
S.~Janowski, D.~Parganlija, F.~Giacosa and D.~H.~Rischke,
Phys.\ Rev.\ D {\bf 84}, 054007 (2011)  [arXiv:1103.3238 [hep-ph]];  
S.~Janowski, F.~Giacosa and D.~H.~Rischke,
Phys.\ Rev.\ D {\bf 90}, no. 11, 114005 (2014)  [arXiv:1408.4921 [hep-ph]].  



\bibitem{Parganlija:2010fz} 
D.~Parganlija, F.~Giacosa and D.~H.~Rischke,
Phys.\ Rev.\ D {\bf 82}, 054024 (2010)  [arXiv:1003.4934 [hep-ph]];  
D.~Parganlija, P.~Kovacs, G.~Wolf, F.~Giacosa and D.~H.~Rischke,
Phys.\ Rev.\ D {\bf 87}, 014011 (2013)  [arXiv:1208.0585 [hep-ph]].  
  
  
\bibitem{Jaffe:1976ig} 
  R.~L.~Jaffe,
Phys.\ Rev.\ D {\bf 15}, 267 (1977);  
A.~H.~Fariborz, R.~Jora and J.~Schechter,
Phys.\ Rev.\ D \textbf{72}, 034001 (2005) [arXiv:hep-ph/0506170];
F.~Giacosa,
  Phys.\ Rev.\ D {\bf 75}, 054007 (2007)
  [hep-ph/0611388];
  A.~Heinz, S.~Struber, F.~Giacosa and D.~H.~Rischke,
Phys.\ Rev.\ D {\bf 79}, 037502 (2009)  [arXiv:0805.1134 [hep-ph]];  
  A.~Heinz, S.~Struber, F.~Giacosa and D.~H.~Rischke,
Acta Phys.\ Polon.\ Supp.\  {\bf 3}, 925 (2010)  [arXiv:1006.5393 [hep-ph]].  


\bibitem{Bugg:2004xu} 
D.~V.~Bugg,
Phys.\ Rept.\  {\bf 397}, 257 (2004)  [hep-ex/0412045];  
E.~Klempt and A.~Zaitsev,
Phys.\ Rept.\  {\bf 454}, 1 (2007)  [arXiv:0708.4016 [hep-ph]].  

\bibitem{Parganlija:2012nc} 
D.~Parganlija,
J.\ Phys.\ Conf.\ Ser.\  {\bf 426}, 012019 (2013)  [arXiv:1211.4804 [hep-ph]];  
D.~Parganlija,
J.\ Phys.\ Conf.\ Ser.\  {\bf 503}, 012010 (2014)  [arXiv:1312.2830 [hep-ph]].  
  

\bibitem{Maldacena} 
  J.~M.~Maldacena,
Int.\ J.\ Theor.\ Phys.\  {\bf 38}, 1113 (1999)  [Adv.\ Theor.\ Math.\ Phys.\  {\bf 2}, 231 (1998)]  [hep-th/9711200].  


\bibitem{Witten} 
  E.~Witten,
Adv.\ Theor.\ Math.\ Phys.\  {\bf 2}, 505 (1998)  [hep-th/9803131].  

\bibitem{Klebanov2000} 
  I.~R.~Klebanov and M.~J.~Strassler,
  JHEP {\bf 0008}, 052 (2000)
  [hep-th/0007191];
  J.~Babington, J.~Erdmenger, N.~J.~Evans, Z.~Guralnik and I.~Kirsch,
  Phys.\ Rev.\ D {\bf 69}, 066007 (2004)
  [hep-th/0306018];
  M.~Kruczenski, D.~Mateos, R.~C.~Myers and D.~J.~Winters,
  JHEP {\bf 0405}, 041 (2004)
  [hep-th/0311270].


\bibitem{SaSu} 
T.~Sakai and S.~Sugimoto,
Prog.\ Theor.\ Phys.\ \textbf{113}, 843 (2005) [hep-th/0412141];
T.~Sakai and S.~Sugimoto,
Prog.\ Theor.\ Phys.\ \textbf{114}, 1083 (2005) [hep-th/0507073];
T.~Imoto, T.~Sakai and S.~Sugimoto,
  Prog.\ Theor.\ Phys.\  {\bf 124}, 263 (2010)
  [arXiv:1005.0655 [hep-th]];
F.~Br\"{u}nner, D.~Parganlija and A.~Rebhan,
Acta Phys.\ Polon.\ Supp.\  {\bf 7}, no. 3, 533 (2014)  [arXiv:1407.6914 [hep-ph]];  
 A.~Rebhan,
arXiv:1410.8858 [hep-th].  



  
\bibitem{Afonin:2010fr} 
  S.~S.~Afonin,
  Int.\ J.\ Mod.\ Phys.\ A {\bf 25}, 5683 (2010)
  [arXiv:1001.3105 [hep-ph]].

\bibitem{BPR} 
  F.~Br\"{u}nner, D.~Parganlija and A.~Rebhan,
  arXiv:1501.07906 [hep-ph].

  
  
  
\bibitem{BMT} 
  R.~C.~Brower, S.~D.~Mathur and C.~I.~Tan,
Nucl.\ Phys.\ B {\bf 587}, 249 (2000)  [hep-th/0003115].  
  
\bibitem{CM} 
  N.~R.~Constable and R.~C.~Myers,
JHEP {\bf 9910}, 037 (1999)  [hep-th/9908175].  





\bibitem{Hashimoto} 
  K.~Hashimoto, C.~-ITan and S.~Terashima,
  Phys.\ Rev.\ D {\bf 77}, 086001 (2008)
  [arXiv:0709.2208 [hep-th]].

  
\bibitem{PhD} 
  D.~Parganlija,
arXiv:1208.0204 [hep-ph].  



\bibitem{PANDA} 
  M.~F.~M.~Lutz {\it et al.}  [PANDA Collaboration],
  arXiv:0903.3905 [hep-ex].
  
  

  
\end{thebibliography}
\end{document}